\documentclass[12pt]{amsart}

\usepackage[dvips]{graphics}

\usepackage{epsfig}

\begin{document}

\title[]{\bf Perturbative Evaluation of Interacting Scalar
             Fields on a Curved Manifold
             with Boundary}

\author[]{George Tsoupros \\
       {\em The School of Physics,}\\
       {\em Peking University,}\\
       {\em Beijing 100871,}\\
       {\em People's Republic of China}
}
\thanks{present e-mail address: gts@pku.edu.cn}

\begin{abstract}

The effects of quantum corrections to a conformally invariant scalar field theory on a curved manifold of positive constant curvature with boundary are considered in the context of a renormalisation procedure. The renormalisation of the theory to second order in the scalar self-coupling pursued herein involves explicit calculations of up to third loop-order and reveals that, in addition to the renormalisation of the scalar self-coupling and scalar field, the removal of all divergences necessitates the introduction of conformally non-invariant counterterms proportional to $ R\Phi^2$ and $ K\Phi^2$ in the bare scalar action as well as counterterms proportional to $ RK^2$, $ R^2$ and $ RK$ in the gravitational action. The substantial backreaction effects and their relevance to the renormalisation procedure are analysed.       


\end{abstract}

\maketitle


{\bf I. Introduction}\\

The development of Euclidean Quantum Gravity rendered the physical significance of manifolds with boundary
manifest. That significance has been substantially enhanced by the peculiar implications of the holographic
principle and the $ AdS/CFT$ correspondence. For that matter, issues related to the number of degrees of 
freedom and to the dynamical behaviour of physical systems accommodated on bounded manifolds at fundamental 
level are of immediate importance. The present work addresses the issue of  renormalisation of a conformal 
scalar field theory on $ C_4$, the four-dimensional Riemannian manifold of positive constant curvature bounded 
by a hypersurface of positive constant extrinsic curvature. Such a theory is of direct relevance to Euclidean 
Quantum Cosmology and its renormalisation above one-loop level necessitates the results already attained in 
the context of the technique of the method of images \cite{G}, \cite{T}, \cite{GT}. The stated technique 
exploits the symmetry of the theory in order to relate the eigenvalue problem of the bounded Laplace operator 
on the $ n$-dimensional spherical cap $ C_n$ to that of the unbounded Laplace operator on the covering 
manifold which the $ n$-sphere $ S_n$ constitues. The immediate advantage of such a technique is the 
evaluation of diagrams at any loop-order. 

The renormalisation of the stated theory will be pursued to second order in the scalar self-coupling. Such a 
renormalisation procedure involves calculations of up to third loop-order. It will be shown that, at higher 
loop-orders, the theory manifests a rich structure in the context of which the presence of the boundary substantially enhances 
the dynamical behaviour of the quantum field. In addition to the already established results of simultaneous renormalisation of surface and volume terms and to the interaction between them on a general manifold with boundary \cite{G}, \cite{T}, \cite{GT}, \cite{Od}, \cite{Odintsov}, \cite{Solodukhin} the theory exhibits an unusual 
local dependence. It will be shown that, unlike the situation on unbounded manifolds, the radiative contributions at two-loop level to the bare parameters of the theory are sensitive to the direction of the boundary. Such a situation is, necessarily, the result of the gravitational backreaction on the geometry of positive constant curvature at length
scales relevant to two-loop vacuum effects. The renormalisation procedure reveals that already at scales relevant to two-loop vacuum contributions the semi-classical approximation specified by a quantum field fluctuating on the fixed geometrical background of $ C_4$ breaks down and the semi-classical geometry itself deviates from the fixed geometry of the gravitational instanton as a result of the gravitational backreaction. 
As a consequence, in the embedding 
Euclidean space the residues associated with the two-point function in the context of dimensional regularisation manifest a dependence on the angle specifying the direction
of the boundary. For that matter, the bare parameters relevant to the 
two-point function manifest an unusual time-dependence, albeit in imaginary time. 

In the context of the semi-classical approach to Quantum Gravity any matter propagator in a general curved space-time receives contributions to all orders in the metric tensor from actual gravitons treated as linearised perturbations of the
gravitational field and propagating on the fixed geometrical background which describes space-time \cite{Birrel}. The merit of the spherical formulation of massless scalar theories - which the present diagramatic technique exploits through the method of images - lies in the closed form in which the sum over all graviton contributions results for the scalar propagator \cite{Drummond}, \cite{I.Drummond}, \cite{DrummondShore}, \cite{Shore}, \cite{G.Shore}. In revealing the contributions which the background geometry receives as a result of scalar vacuum effects the evaluation of higher loop-order diagramatic structures signals the associated length scales at which the validity of the spherical formulation itself terminates.

The immediate issue which naturally emerges in such a situation is that of specifying the 
exact relation between the backreaction effects and the renormalisation of the scalar field on $ C_4$. It will be shown that, on condition of the stated space-time dependence, 
the scalar vacuum fluctuations on the modified semi-classical background have a direct contribution to the perturbative redefinition of scalar amplitudes and physical parameters on, what is at larger scales, the geometry of positive constant curvature of $ C_4$.  
As the backreaction-related fluctuations of the semi-classical geometry are relevant exclusively to the high-energy limit of the renormalisation group behaviour of the quantum scalar field the radiative contributions relevant to the renormalisation of the theory to second order in the scalar self-coupling will be pursued in what follows on the fixed semi-classical geometrical background of positive constant curvature until such backreaction effects on the geometry become evident. The additional redefinitions which the scalar vacuum effects signify in the context of gravitational backreaction will be consistently pursued only in their relation to the renormalisation procedure on $ C_4$.

The mathematical expressions which emerge in the process of the ensuing analysis are frequently involved. This feature is expected in view of the small symmetry which the theory accommodates in comparison to the Euclidean de Sitter invariance $ SO(5)$ on $ S_4$. The pole structures associated with each sector in the bare action, however, will be 
obvious. The major results in \cite{G}, \cite{T} and \cite{GT} necessary for the renormalisation procedure are cited in the Appendix a reading of which is recommended before studying the analysis in the ensuing sections.

{\bf II. Scalar Field Renormalisation on $ C_4$}\\ 

The unique conformally invariant scalar action on a n-dimensional Riemannian 
manifold of positive constant embedding radius $ \it{a}$ is \cite{McKeon Tsoupros}

\begin{equation}
S[\Phi] = \int_{S}d^n\eta\big{[}\frac{1}{2}\Phi(\frac{L^2-\frac{1}{2}n(n-2)}{2a^2})\Phi - 
\frac{\lambda}{\Gamma(p+1)}\Phi^p]\big{]}
\end{equation}
with $ p = \frac{2n}{n-2}, n>2$ with $ \eta$ being the position vector in the embedding 
$ n + 1$ Euclidean space signifying the coordinates $ \eta_{\mu}$ and with the generator of rotations being 

$$
L_{\mu \nu} = \eta_{\mu}\frac{\partial}{\partial \eta_{\nu}} -
\eta_{\nu}\frac{\partial}{\partial \eta_{\mu}} 
$$
on the relevant embedded manifold. In the physically relevant case of $ n=4$ the 
self-coupling is that of $ \Phi^4$. Any n-dimensional Riemannian manifold of positive 
constant embedding radius $ \it{a}$ represents, through the usual rotation in imaginary 
time, the Euclidean version of the associated segment of de Sitter space. On any such 
manifold the Ricci scalar $ R$ admits the constant value \cite{McKeon Tsoupros}

\begin{equation}
R = \frac{n(n-1)}{a^2}
\end{equation}
In conformity with the considerations hitherto outlined the Riemannian manifold relevant 
to $ S[\Phi]$ will be specified to be that of a spherical n-cap $ C_n$ considered as a 
manifold of positive constant curvature embedded in a $ (n+1)$-dimensional Euclidean space and 
bounded by a $ (n-1)$-sphere of positive constant extrinsic curvature $ K$ (diverging normals). With respect to $ \eta_{\mu}$ expressed in spherical polar coordinates such a boundary is specified by the angle $ \theta_{n}^0$.   
In effect, $ \it{C_n}$ is characterised by spherical $(n-1)$-dimensional sections of 
constant Euclidean time $ \tau$ or, equivalently, constant extrinsic curvature $ k$. Allowing for surface terms, $ S[\Phi]$ is also the semi-classical action for the conformal scalar field on the bounded segment of $ S_n$ which constitutes the $ n$-dimensional spherical cup $ C_n$. Semiclassically, the condition stipulated on the boundary will be the Dirichlet condition $ \Phi_{\partial C} = 0$. Since, in principle, all conformally  
non-invariant counterterms are perturbatively possible as a result of radiative effects the bare scalar action on 
$ C_n$ defining a renormalisable scalar theory at $ n=4$ will be

\newpage

$$
S[\Phi_0] = \int_{C}d^n\eta\big{[}\frac{1}{2}\Phi_0(\frac{L^2-\frac{1}{2}n(n-2)}{2a^2})\Phi_0 - \frac{\lambda_0}{4!}\Phi_0^4 - \frac{1}{2}[m_0^2 + \xi_0 R + \kappa_0K^2]\Phi_0^2 \big{]} 
$$

\begin{equation}
+ \oint_{\partial C}d^3\eta(\epsilon_0K\Phi_0^2)
\end{equation} 
with $ R$ and $ K$ being the constant scalar curvature and constant extrinsic curvature of the boundary hypersurface respectively, with the semi-classical Gibbons-Hawking surface term for a conformal scalar field considered as a component of the matter rather than the gravitational action \cite{G} and on the understanding that all three conformally non-invariant sectors in the volume of $ C_n$ are absent at the semi-classical limit. In principle all three such sectors are expected as a result of radiative contributions to the two-point function.

The scalar action in (3) is only the matter component of the bare action. The scalar theory defined exclusively by (3) is not renormalisable. The gravitational component of the bare action on $ C_n$ at $n \rightarrow 4$ is

\begin{equation}
S_g = \int_Cd^4\eta\big{[}\frac{1}{8\pi G_0}\Lambda - \frac{1}{16\pi G_0}(R + K^2) + \alpha_0R^2 + \beta_0RK^2\big{]}
+ \oint_{\partial C}d^3\eta\big{[}\gamma_0RK + \delta_0K^3\big{]}
\end{equation}   
Attention is invited to the fact that the underlying spherical formulation of massless scalar field theories
for which the method of images allows determines all Green functions as an exact function of the scalar curvature $ R$.
On the bounded spherical cap $ C_n$ such an evaluation renders the Green functions also  exact functions of $ K$.
This feature of the formalism which has both perturbative and formal significance \cite{Drummond}, \cite{McKeon Tsoupros} accounts for the absence in $ S_g$ of the sectors relevant to the Ricci and Riemann tensor-related contractions $ R_{\mu\nu}R^{\mu\nu}$ and $ R_{\alpha\beta\gamma\delta}R^{\alpha\beta\gamma\delta}$ respectively.
Effectively, only volume-related counterterms proportional to the cosmological constant $ \Lambda$ as well as to the dimensionally consistent powers of $ R, K$ and their multiplicative combinations featured in $ S_g$ are allowed on 
$ C_4$. This is the case because, perturbatively, the only source of volume divergences are terms in the relevant expansions of the Green functions.

The bare action $ S$ of the theory for $ n=4$ in the context of multiplicative renormalisation is 

\begin{equation}
S = S_g + S[\Phi_0]   
\end{equation}
With the obvious exceptions of the bare gravitational constant $ G_0$ and the square of the bare mass $ m_0$ all other 
couplings in $ S$ are characterised by the absence of mass dimensions at $ n=4$. 
Moreover, the interaction between surface and volume terms \cite{G}, \cite{Solodukhin} renders, effectively, the extrinsic curvature $ K$ of the manifold's boundary as yet another coupling in the theory. 
In the context of both dimensional regularisation - which manifests all divergences as poles in $ \epsilon = 4-n$ at the dimensional limit $ n \rightarrow 4$ after an analytical extension of $ n$ - and minimal subtraction - which is characterised by the absence of finite parts in the counterterms at $ n \rightarrow 4$ - the expansions of the bare parameters in the scalar action in powers of the poles and of the scalar self-coupling $ \lambda$ \cite{'t Hooft} are 

$$
(6a)\hspace{1.0in}
\lambda_0 = \mu^{\epsilon}\big{[}\lambda + \sum_{\nu=1}^{\infty}\frac{a_{\nu}(\lambda)}{\epsilon^{\nu}}\big{]} = 
\mu^{2\epsilon}\big{[}\lambda + \sum_{k=1}^{\infty}\sum_{i=k}^{\infty}\frac{a_{ki}\lambda^i}{\epsilon^{k}}\big{]}
\hspace{2.5in} 
$$

$$
(6b)\hspace{1.0in}
m_0^2 = m^2\big{[}1 + \sum_{\nu=1}^{\infty}\frac{b_{\nu}^{(m^2)}(\lambda)}{\epsilon^{\nu}}\big{]} +
mK\sum_{\nu=1}^{\infty}\frac{b_{\nu}^{(m)}(\lambda)}{\epsilon^{\nu}}
\hspace{2.5in}
$$

$$
(6c)\hspace{1.0in}
R\xi_0 = R\big{[}\frac{n-2}{4(n-1)} + \sum_{\nu=1}^{\infty}\frac{c_{\nu}(\lambda)}{\epsilon^{\nu}}\big{]} = 
R\big{[}\frac{n-2}{4(n-1)} + \sum_{k=1}^{\infty}\sum_{i=k}^{\infty}\frac{c_{ki}\lambda^i}{\epsilon^{k}}\big{]}
\hspace{2.5in} 
$$

$$
(6d)\hspace{1.0in}
K^2\kappa_0 = K^2\sum_{\nu=1}^{\infty}\frac{d_{\nu}(\lambda)}{\epsilon^{\nu}} = 
K^2\sum_{k=1}^{\infty}\sum_{i=k}^{\infty}\frac{d_{ki}\lambda^i}{\epsilon^{k}}
\hspace{2.5in} 
$$

$$
(6e)\hspace{1.0in}
K\epsilon_0 = K\sum_{\nu=1}^{\infty}\frac{e_{\nu}(\lambda)}{\epsilon^{\nu}} =
K\sum_{k=1}^{\infty}\sum_{i=k}^{\infty}\frac{e_{ki}\lambda^i}{\epsilon^{k}}
\hspace{2.5in}    
$$

$$
(6f)\hspace{1.0in}
Z = 1 + \sum_{\nu=1}^{\infty}\frac{f_{\nu}(\lambda)}{\epsilon^{\nu}} = 
1+ \sum_{k=1}^{\infty}\sum_{i=k}^{\infty}\frac{f_{ki}\lambda^i}{\epsilon^{k}} 
\hspace{2.5in}    
$$
whereas the expansions of the bare parameters of the gravitational action will be

$$
(7a)\hspace{1.0in}
\alpha_0 = \sum_{\nu=1}^{\infty}\frac{\alpha_{\nu}(\lambda)}{\epsilon^{\nu}} = 
\sum_{k=1}^{\infty}\sum_{i=k}^{\infty}\frac{\alpha_{ki}\lambda^i}{\epsilon^{k}} 
\hspace{2.5in}    
$$

$$
(7b)\hspace{1.0in}
\beta_0 = \sum_{\nu=1}^{\infty}\frac{\beta_{\nu}(\lambda)}{\epsilon^{\nu}} = 
\sum_{k=1}^{\infty}\sum_{i=k}^{\infty}\frac{\beta_{ki}\lambda^i}{\epsilon^{k}} 
\hspace{2.5in}    
$$

$$
(7c)\hspace{1.0in}
\gamma_0 = \sum_{\nu=1}^{\infty}\frac{\gamma_{\nu}(\lambda)}{\epsilon^{\nu}} = 
\sum_{k=1}^{\infty}\sum_{i=k}^{\infty}\frac{\gamma_{ki}\lambda^i}{\epsilon^{k}} 
\hspace{2.5in}    
$$

The semi-classical mass parameter m is, generally, a free parameter. In view of the absence of mass at
semi-classical level in the present theory the parameter m refers, evidently, to the order at which the 
mass is possibly generated through vacuum effects. It will be shown that to $ O(\lambda^2)$ there is no
mass generation in the theory and the speculation herein is that it is absent to all orders. As stated, 
in what follows the regulating scheme will be predicated on an analytical extension of the covering 
manifold's dimensionality to an arbitrary value $ n$. Such an approach is preferable as it preserves the 
maximal symmetry which underlies the method of images. However, the formal extension of $ S_4$ to 
topologically flat dimensions such as $ S_4 \times R_{n-4}$ as opposed to $ S_n$ will have no impact on
the renormalisation group behaviour of the theory and it is expected, in addition to leave its 
pole-structure intact \cite{GT}. In addition, consistency with the renormalisation group behaviour of 
the theory necessitates that the coupling in the conformal sector at semi-classical level be taken, in 
(1), (3) and (6c), at its $ n$-dimensional value of $ \xi(n)=\frac{1}{4}\frac{n-2}{n-1}$ instead of at its 
four-dimensional value of $ \frac{1}{6}$ \cite{GT}.

Pursuant to the spherical formulation relevant to diagramatic evaluations on $ C_n$ developed in \cite{T} all calculations will be advanced in configuration space. In the context of the Dirichlet condition $ \Phi_{\partial C}=0$ the  bounded operator 

$$ M \equiv \frac{L^2-\frac{1}{2}n(n-2)}{2a^2}$$ 
on $ C_n$ which appears in (3) has been shown to admit the Green function \cite{G}   

\addtocounter{equation}{2}%
\begin{equation}
D_{c}^{(n)}(\eta,{\eta}') = \frac{\Gamma(\frac{n}{2}-1)}{4\pi^{\frac{n}{2}}}\frac{1}{|{\eta}-{\eta}'|^{n-2}} 
- \frac{\Gamma(\frac{n}{2}-1)}{4\pi^{\frac{n}{2}}}\frac{1}{|\frac{a_{{\eta}'}}{a_B}{\eta}-
\frac{a_B}{a_{{\eta}'}}{\eta}'|^{n-2}}
\end{equation}

satisfying 

\begin{equation}
MD_{c}^{(n)}(\eta,{\eta}') = -\delta^{(4)}(\eta - \eta')
\end{equation}
The parameters $ a_{\eta'}$ and $ a_{B}$ in (8) are the geodesic distances on $ C_n$ between its pole and point $ \eta'$ as well as between its pole and the boundary. The singular part in (8) is identical to the propagator on $ S_n$ whereas the boundary part remains always finite and ensures the absence of propagation on $ \partial C_n$.

The diagrams relevant to the ensuing renormalisation procedure to second order in the scalar self-coupling $ \lambda$ are cited in the following figure

\begin{figure}[h]
\centering\epsfig{figure=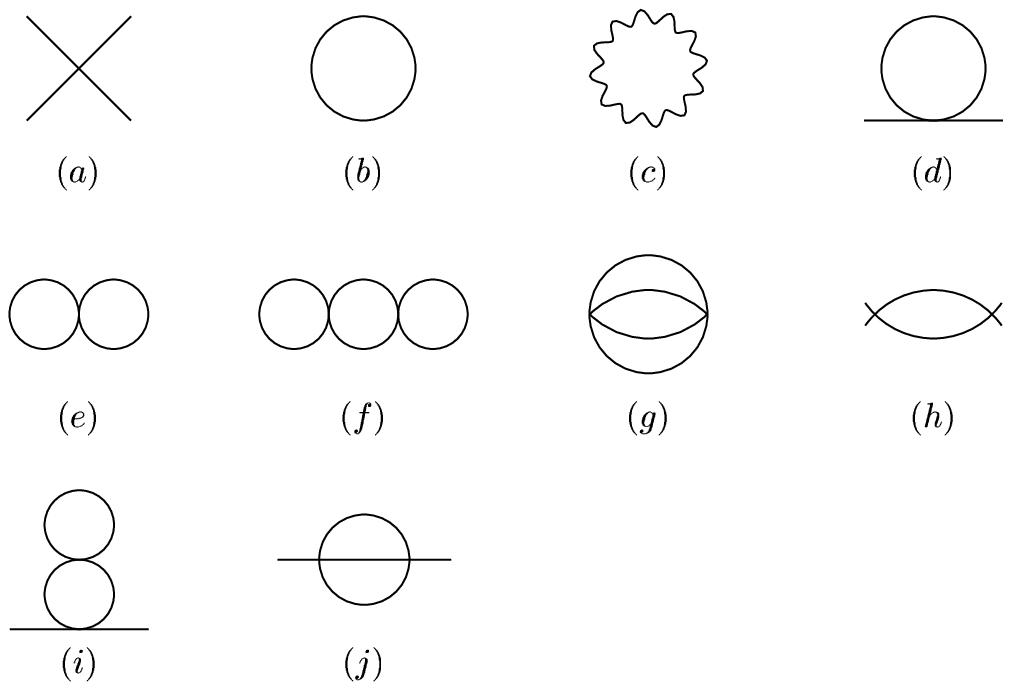, height = 70mm,width=100mm}
\caption{Diagrams contributing to $ O(\lambda^2)$-related renormalisation.}
\end{figure}
The ``bubble'' diagrams representing scalar and graviton loops in fig.(1b) and fig.(1c) respectively account for the one-loop contribution to the zero-point function of the theory. Their simultaneous presence in any curved space-time is expected on the basis of general theoretical considerations \cite{Birrel}. They are characterised by the absence of interaction vertices and, on power counting grounds, are responsible for the simultaneous one-loop contributions to volume and boundary effective Einstein-Hilbert action on any manifold with boundary \cite{Solodukhin}. They have been shown to be finite provided that dimensional regularisation is used \cite{G}. Diagram (1a) represents the zero-loop (tree level) contribution to the four-point function. In the context of multiplicative renormalisation it receives its expression 

\begin{equation}
(1a): -\lambda_0
\end{equation}
from the scalar self-interacting sector in (3). Diagram (1d) is finite to first order in $ \lambda$. However, since to $ O(\lambda^2)$ this diagram contains an overlapping divergence the result of its evaluation in \cite{G} is reproduced in (A3'). Diagrams (1e), (1f) and (1g) represent vacuum contributions to the zero-point function and are relevant to renormalisation in the gravitational action. Diagram (1g) has been evaluated in \cite{GT}. The result is reproduced in (A5). The mathematical expression for the diagramatic structure of (1e) is obtained directly from that of that of (1d) through (A3). With the associated symmetry factor of $ \frac{1}{4}$ this expression in $ n$-dimensions is

$$
(1e)_n: 
(-\lambda_0)\frac{1}{4}\big{[}\frac{\Gamma(\frac{n}{2}-1)}{4\pi^{\frac{n}{2}}}\big{]}^2\int_Cd^n\eta\big{[}-|\frac{a_{\eta}}{a_B}\eta - \frac{a_{B}}{a_\eta}\eta|^{2-n}\big{]}^2  
$$
which, in the same calculational context that resulted in (A3'), yields

$$
(1e)_n: 
(-\lambda_0)\frac{1}{4}\big{[}\frac{\Gamma(\frac{n}{2}-1)}{4\pi^{\frac{n}{2}}}\big{]}^2\frac{1}{16\pi^n}\big{[}\frac{1}{(n-2)!}\big{]}^2\big{[}\Gamma(\frac{n-1}{2})\big{]}^2\times
$$

$$
a^{4-2n}\big{[}\sum_{N=0}^{N_0}\frac{(2N+n-1)\Gamma(N+n-1)}{(N+\frac{n}{2})(N+\frac{n}{2} - 1)\Gamma(N + 1)}\big{]}^2
\int_Cd^n\eta   
$$

eventuating at $ n \rightarrow 4$ in the result

\begin{equation}
(1e): 
(-\lambda_0)\frac{1}{3^2}\frac{1}{2^{19}}\frac{V_c + 1}{\pi^7}\big{[}\sum_{N=0}^{N_0}(2N + 3)\big{]}^2 R^2\int_Cd^n\eta 
\end{equation}
where use has also been made of (2).

Diagram (1f) features two vertices at $ \eta$ and $ \eta'$ in addition to the massless "tadpole" loop-structures expressed by (A3). With the symmetry factor of $ \frac{1}{16}$ its mathematical expression is

$$
(1f)_n: (-\lambda_0)^2\frac{1}{16}\big{[}\frac{\Gamma(\frac{n}{2}-1)}{4\pi^{\frac{n}{2}}}\big{]}^2 \times
$$

$$
\int_Cd^n\eta\int_Cd^n\eta'\big{[}-|\frac{a_{\eta'}}{a_B}\eta' - \frac{a_{B}}{a_\eta'}\eta'|^{2-n}\big{]}\big{[}-|\frac{a_{\eta}}{a_B}\eta - \frac{a_{B}}{a_\eta}\eta|^{2-n}\big{]}
\big{[}D_c(\eta - \eta')\big{]}^2
$$
In the context of (A1), (A2) and (8) it is obvious that only the term featuring the square of the singular part in the expansion of the propagator $ [D_c(\eta - \eta')]^2$ yields a divergent result. For that matter

$$
(1f)_n: (-\lambda_0)^2\frac{1}{16}\big{[}\frac{\Gamma(\frac{n}{2}-1)}{4\pi^{\frac{n}{2}}}\big{]}^2 \times
$$

$$
\int_Cd^n\eta\int_Cd^n\eta'\big{[}-|\frac{a_{\eta'}}{a_B}\eta' - \frac{a_{B}}{a_\eta'}\eta'|^{2-n}\big{]}\big{[}-|\frac{a_{\eta}}{a_B}\eta - \frac{a_{B}}{a_\eta}\eta|^{2-n}\big{]}
|\eta - \eta'|^{2(2-n)} + F. T.
$$
The diagramatic structures which feature powers of the singular or boundary part of the propagator are generally evaluated, through (A1) and (A2), in the same context as that relevant to the derivation of the expression for diagram (1g) in (A5) \cite{GT}. The procedure is outlined in the context of the present diagram in question. It is

$$
(1f)_n:
(-\lambda_0)^2\frac{1}{16}\big{[}\frac{\Gamma(\frac{n}{2}-1)}{4\pi^{\frac{n}{2}}}\big{]}^4\frac{1}{16\pi^n}\big{[}\frac{1}{(n-2)!}\big{]}^2\big{[}\Gamma(\frac{n-1}{2})\big{]}^2 a^{4-2n}\times
$$

$$
\big{[}\sum_{N'=0}^{N'_0}\frac{(2N'+n-1)\Gamma(N'+n-1)}{(N'+\frac{n}{2})(N'+\frac{n}{2} - 1)\Gamma(N' + 1)}\big{]}^2 
\sum_{N=0}^{\infty}\sum_{\alpha=0}^{N}
\frac{(2a)^{4-n}{\pi}^{\frac{n}{2}}\Gamma(2-\frac{n}{2})\Gamma(N-2+n)}{
\Gamma(N+2)\Gamma(n-2)}\times$$

$$
\int_Cd^n\eta Y_{\alpha}^N({\eta})\int_Cd^n\eta' Y_{\alpha}^N({\eta}')
$$

The integrals in this expression are expressed \cite{GT} through use of 

\begin{equation}
Y_{0}^{0} = \frac{1}{\sqrt{a^n \Omega_{n+1}}}
\end{equation} 
in the context of (A9) and (A10) as

$$
\int_Cd^n\eta Y_{\alpha}^N({\eta})\int_Cd^n\eta' Y_{\alpha}^N({\eta}') = {a^n \Omega_{n+1}} 
\int_Cd^n\eta Y_{\alpha}^N(\eta)Y_0^0\int_Cd^n\eta' Y_{\alpha}^N(\eta')Y_0^0 = 
$$

\begin{equation}
{a^n \Omega_{n+1}}\frac{a^2K^2}{[(\frac{n}{2}-1)\frac{n}{2} - (N+\frac{n}{2}-1)(N+\frac{n}{2})]^2}[C_N^{\frac{n-1}{2}}(cos{\theta_n^0})]^2
\end{equation}
where use has been made of \cite{Bateman} 

\begin{equation}
C_0^k(x) = 1 ; k > -\frac{1}{2}
\end{equation}
Further, in view of the simple pole in $ \epsilon = 4-n$ which $ \Gamma(2-\frac{n}{2})$ generates at the dimensional limit $ n \rightarrow 4$ it is

\newpage

$$
(1f)_n \rightarrow
(-\lambda_0)^2\frac{1}{2^{14}}\frac{1}{\pi^2}[\Gamma(\frac{3}{2})]^2a^{-2}(a^4\Omega_5)K^2\times $$

$$
\big{[}\sum_{N'=0}^{N'_0}(2N'+3)\big{]}^2 
\sum_{N=1}^{\infty}\frac{[C_N^{\frac{3}{2}}(cos{\theta_4^0})]^2}{N^2(N+3)^2}
\big{[}\frac{2}{\epsilon} - \gamma_E + \frac{\frac{\pi^2}{6}+\gamma_E^2}{4}\epsilon + ....\big{]}
$$
It is reminded that inherent in (A11) is the condition $ N \neq N'$ which accounts for the absence of an infra-red divergence. The limit of any power of $ C_N^k(x)$ over the degree $ N$ for a fixed order $ k$ has a finite radius of convergence between $ (-1,1)$ \cite{Bateman}. For that matter, the series over $ N$ in the expression above is convergent. A further use of (2), (A4), (A4') as well as of

$$ 
\Gamma(1 + z) = z\Gamma(z)~~~; ~~~z \in C $$ 

finally yields 

$$
(15)\hspace{0.4in} (1f): 
(-\lambda_0)^2\frac{1}{3}\frac{1}{2^{17}}\frac{V_c+1}{\pi}\sum_{N=1}^{\infty}\frac{[C_N^{\frac{3}{2}}(cos{\theta_4^0})]^2}{N^2(N+3)^2}\big{[}\sum_{N'=0}^{N'_0}(2N'+3)\big{]}^2K^2R \frac{1}{\epsilon} \int_Cd^n\eta + F. T.
\hspace{0.5in}
$$

Diagram (1i) involves two self-interaction vertices and, as is the case with diagram (1d),
represents contributions to the two-point function. Its evaluation advances on the lines of that for diagram (1f). With a symmetry factor of $ \frac{1}{4}$ it is

\addtocounter{equation}{1}%
\begin{equation}
(1i):~~
(-\lambda_0)^2\frac{1}{4}\int_Cd^4\eta \int_Cd^4\eta' D_c(\eta_1 - \eta)[D_c(\eta - \eta')]^2D_c(\eta' - \eta')D_c(\eta - \eta_2)
\end{equation}
Again, only the term featuring the square of the singular part in the expansion of $ [D_c(\eta - \eta')]^2$ yields a divergent result. For that matter, the loop structure of diagram (1i) in n-dimensions is

$$
(-\lambda_0)^2\frac{1}{4}\big{[}\frac{\Gamma(\frac{n}{2}-1)}{4\pi^{\frac{n}{2}}}\big{]}^3 
\int_Cd^n\eta\int_Cd^n\eta'\big{[}-|\frac{a_{\eta'}}{a_B}\eta' - \frac{a_{B}}{a_\eta'}\eta'|^{2-n}\big{]}
|\eta - \eta'|^{2(2-n)} + F. T.
$$
Likewise, use of the expression for the loop-structure of diagram (1d) in (A3') as well as of (A1) yields

$$
-(\lambda_0)^2\big{[}\frac{\Gamma(\frac{n}{2}-1)}{4\pi^{\frac{n}{2}}}\big{]}^3\frac{1}{2^{4}\pi^\frac{n}{2}}\frac{1}{(n-2)!}\Gamma(\frac{n-1}{2})a^{2-n}\times
$$

$$
\sum_{N'=0}^{N'_0}\frac{(2N'+n-1)\Gamma(N'+n-1)}{(N'+\frac{n}{2})(N'+\frac{n}{2} - 1)\Gamma(N' + 1)} 
\sum_{N=0}^{\infty}\sum_{\alpha=0}^{N}
\frac{(2a)^{4-n}{\pi}^{\frac{n}{2}}\Gamma(2-\frac{n}{2})\Gamma(N-2+n)}{
\Gamma(N+2)\Gamma(n-2)}\times$$

$$
\int_Cd^n\eta Y_{\alpha}^N({\eta})\int_Cd^n\eta' Y_{\alpha}^N({\eta}')
$$
Inserting this result in (16) and advancing the calculation on the lines of the derivation of (15) yields the mathematical expression for diagram (1i) as

$$
(1i):\hspace{0.1in}
-(\lambda_0)^2\frac{1}{2^{11}}\frac{1}{\pi^{\frac{11}{2}}}(V_c + 1)\times
$$

\begin{equation}
\sum_{N=1}^{\infty}\frac{[C_N^{\frac{3}{2}}(cos{\theta_4^0})]^2}{N^2(N+3)^2}\sum_{N'=0}^{N'_0}(2N'+3)K^2 \frac{1}{\epsilon} \int_Cd^n\eta D_c(\eta_1 - \eta)D_c(\eta - \eta_2)
+ F. T.
\end{equation}

Diagram (1h) involves two vertices at $ \eta$ and $ \eta'$ and signifies contributions to the four-point function. In view of the three possible channels through which it is realised the associated symmetry factor is $ 3\times\frac{1}{2}$. Its mathematical expression eventuates in a divergence at $ \eta = \eta'$ as $ n \rightarrow 4$. This divergence is contained, for that matter, entirely in the expression 

$$
(1h):~~(-\lambda_0)^2\frac{3}{2}\times
$$

\begin{equation}
\int_Cd^4\eta \int_Cd^4\eta' 
D_c(\eta_1 - \eta)D_c(\eta_2 - \eta)D_c(\eta_3 - \eta)D_c(\eta_4 - \eta)
[D_c(\eta - \eta')]^2
\end{equation}
which is effected by setting $ \eta$ equal to $ \eta'$ in all external propagators. The divergence stems again from the term featuring the square of the singular part of the scalar propagator in the expression for the proper structure of (1h) which, on the same lines as before, yields through use of (A1), the result

$$
\int_Cd^4\eta \int_Cd^4\eta' [D_c(\eta - \eta')]^2 = 
\big{[}\frac{\Gamma(\frac{n}{2}-1)}{4\pi^{\frac{n}{2}}}\big{]}^2 
\int_Cd^n\eta\int_Cd^n\eta'|\eta - \eta'|^{2(2-n)} =
$$

$$
\frac{3}{2\pi^2}(V_c + 1) \sum_{N=1}^{\infty}\frac{[C_N^{\frac{3}{2}}(cos{\theta_4^0})]^2}{N^2(N+3)^2}\frac{K^2}{R} ~~~\frac{1}{\epsilon} \int_Cd^n\eta 
$$

Comparing this expression for the proper diagram with the expression for the connected diagram in (18) reveals that

$$
(1h):~~~(\lambda_0)^2\frac{9}{4\pi^2}(V_c + 1)\sum_{N=1}^{\infty}\frac{[C_N^{\frac{3}{2}}(cos{\theta_4^0})]^2}{N^2(N+3)^2}\frac{K^2}{R} ~~~\frac{1}{\epsilon}\times
$$

\begin{equation}
\int_Cd^n\eta 
D_c(\eta_1 - \eta)D_c(\eta_2 - \eta)D_c(\eta_3 - \eta)D_c(\eta_4 - \eta) + F. T.
\end{equation}

The evaluation of diagram (1h) allows for the renormalisation of the scalar self-coupling.
Indeed, demanding in the context of (6a), that the additive result of (10) and (19) be finite to second order in $ \lambda_0$ yields

\begin{equation}
a_{12}\lambda^2 = \frac{9}{4\pi^2}(V_c + 1)\sum_{N=1}^{\infty}\frac{[C_N^{\frac{3}{2}}(cos{\theta_4^0})]^2}{N^2(N+3)^2}\frac{K^2}{R}
\lambda^2
\end{equation}
In the context of this redefinition all overlapping divergences to $ O(\lambda^2)$ must be rendered manifest. It is obvious by inspection that such divergences are contained exclusively in diagrams (1d) and (1e). The mathematical expression of (1d) becomes through (A3') and (20)

$$
(1d)':\hspace{0.5in}
\mu^{2\epsilon}\big{(}\frac{9}{4\pi^2}(V_c + 1)\sum_{N=1}^{\infty}\frac{[C_N^{\frac{3}{2}}(cos{\theta_4^0})]^2}{N^2(N+3)^2}\frac{K^2}{R}
\lambda^2\big{)}\times
$$

$$
\frac{1}{3}\frac{1}{2^{10}}\frac{V_c + 1}{\pi^{\frac{7}{2}}}\big{[}\sum_{N'=0}^{N'_0}(2N' + 3)\big{]}R\frac{1}{\epsilon}\int_Cd^4\eta D_c(\eta_1 - \eta)D_c(\eta - \eta_2)
$$
Likewise that of (1e) becomes through (11) and (20)

$$
(1e)':\hspace{0.2in}
\frac{1}{3^2}\frac{1}{2^{19}}\frac{V_c + 1}{\pi^7}\big{[}\sum_{N'=0}^{N'_0}(2N' + 3)\big{]}^2 R^2\big{(}\frac{9}{4\pi^2}(V_c + 1)\sum_{N=1}^{\infty}\frac{[C_N^{\frac{3}{2}}(cos{\theta_4^0})]^2}{N^2(N+3)^2}\frac{K^2}{R}
\mu^{2{\epsilon}}\lambda^2\big{)}\frac{1}{\epsilon}
\int_Cd^n\eta 
$$ 

With these redefinitions all diagramatic evaluations hitherto attained have been advanced to second order in the scalar self-coupling. For the sake of convenience the relevant results are listed below

$$
(1d):\hspace{0.1in}
\mu^{2\epsilon}\lambda^2\frac{3}{2^{12}}\frac{1}{\pi^\frac{11}{2}}(V_c + 1)^2\sum_{N=1}^{\infty}\frac{[C_N^{\frac{3}{2}}(cos{\theta_4^0})]^2}{N^2(N+3)^2}
\big{[}\sum_{N'=0}^{N'_0}(2N' + 3)\big{]}K^2\frac{1}{\epsilon}\times
$$

\begin{equation}
\int_Cd^4\eta D_c(\eta_1 - \eta)D_c(\eta - \eta_2) + F. T.
\end{equation}

$$
(1e):~~~-\mu^{2{\epsilon}}
\lambda^2\frac{1}{2^{21}}\frac{(V_c + 1)^2}{\pi^9}\big{[}\sum_{N'=0}^{N'_0}(2N' + 3)\big{]}^2
\sum_{N=1}^{\infty}\frac{[C_N^{\frac{3}{2}}(cos{\theta_4^0})]^2}{N^2(N+3)^2} 
RK^2\frac{1}{\epsilon}\times
$$

\begin{equation}
\int_Cd^n\eta + F. T.
\end{equation}

$$
(1f):\hspace{0.3in}
\mu^{2{\epsilon}}\lambda^2\frac{1}{3}\frac{1}{2^{17}}\frac{V_c+1}{\pi}\sum_{N=1}^{\infty}\frac{[C_N^{\frac{3}{2}}(cos{\theta_4^0})]^2}{N^2(N+3)^2}\big{[}\sum_{N'=0}^{N'_0}(2N'+3)\big{]}^2RK^2 \frac{1}{\epsilon}\times 
$$

\begin{equation}
\int_Cd^n\eta + F. T.
\end{equation}

$$
(1i):\hspace{0.1in}
-\mu^{2{\epsilon}}\lambda^2\frac{1}{2^{11}}\frac{1}{\pi^{\frac{11}{2}}}(V_c + 1)
\sum_{N=1}^{\infty}\frac{[C_N^{\frac{3}{2}}(cos{\theta_4^0})]^2}{N^2(N+3)^2}\sum_{N'=0}^{N'_0}(2N'+3)K^2 \frac{1}{\epsilon}\times
$$

\begin{equation}
\int_Cd^n\eta D_c(\eta_1 - \eta)D_c(\eta - \eta_2)
+ F. T.
\end{equation}

$$
(1h):\hspace{0.7in}\mu^{2{\epsilon}}\lambda^2\frac{9}{4\pi^2}(V_c + 1)\sum_{N=1}^{\infty}\frac{[C_N^{\frac{3}{2}}(cos{\theta_4^0})]^2}{N^2(N+3)^2}\frac{K^2}{R} ~~~\frac{1}{\epsilon}\times
\hspace{0.7in}
$$

\begin{equation}
\int_Cd^n\eta 
D_c(\eta_1 - \eta)D_c(\eta_2 - \eta)D_c(\eta_3 - \eta)D_c(\eta_4 - \eta) + F. T.
\end{equation}
whereas the expression for diagram (1g) is that of (A5) with $ \lambda_0$ replaced by
$ \lambda\mu^{{\epsilon}}$.

The evaluation of diagram (1j) is substantially more involved than that of any other in figure (1). With the external wavefunctions represented by scalar propagation between points $ \eta_1, \eta$ and $ \eta', \eta_2$ respectively it signifies, to second order in $ \lambda_0$, a radiative contribution to the two-point function for the scalar field inherent in the expression

$$
(1j):\hspace{1.5in}
\int_Cd^n\eta\int_Cd^n\eta'D_c(\eta_1 - \eta)\big{[}D_c(\eta - \eta'){]}^3D_c(\eta' - \eta_2)
\hspace{1.5in}
$$
Despite appearances, the external propagators also have a contribution to the divergent structure of this loop-integral. Following an argument in \cite{Drummond} the evaluation of the expression above necessitates an expansion of $ D_c(\eta'-\eta_2)$ about $ \eta'=\eta$ through which it becomes

\begin{equation}
(1j):~~~
\int_Cd^n\eta\int_Cd^n\eta'D_c(\eta_1 - \eta)\big{[}D_c(\eta - \eta')\big{]}^3\big{[}1 + \frac{(\eta - \eta')^2}{2na^2}\frac{L^2}{2}\big{]}D_c(\eta - \eta_2)
\end{equation}
The evaluation of the potential contributions to the divergent structure of (26) necessitate the expansion

$$
\big{[}D_c(\eta - \eta')\big{]}^3 = 
\left[ \frac{\Gamma(\frac{n}{2} - 1)}{4\pi^{\frac{n}{2}}}\right]^3
\big{[} |\eta - \eta'|^{3(2-n)} - 3|\eta - \eta'|^{2(2-n)}|\frac{a_{\eta'}}{a_{B}}\eta - 
\frac{a_{B}}{a_{\eta'}}\eta'|^{2-n} + 
$$

\begin{equation}
3|\eta - \eta'|^{2-n}|\frac{a_{\eta'}}{a_{B}}\eta - \frac{a_{B}}{a_{\eta'}}\eta'|^{2(2-n)} - |\frac{a_{\eta'}}{a_{B}}\eta - 
\frac{a_{B}}{a_{\eta'}}\eta'|^{3(2-n)}\big{]} 
\end{equation}
After replacing (27) in (26) the resulting first term is seen, through use of (A1), to be predicated on the integral

\begin{equation}
\int_Cd^n\eta'|\eta - \eta'|^{3(2-n)} = 
\sum_{N=0}^{\infty}\sum_{\alpha = 0}^N
\frac{(2a)^{6-2n}{\pi}^{\frac{n}{2}}\Gamma(3-n)\Gamma(N-3+\frac{3n}{2})}{
\Gamma(N+3-\frac{n}{2})\Gamma(\frac{3n}{2}-3)}Y_{\alpha}^N({\eta})\int_Cd^n\eta'Y_{\alpha}^N(\eta')
\end{equation}
As was the case with (13) the integral over $ \eta'$ in this expression can be manipulated through (A9), (A10) and (A11) to result in

\newpage

$$
\int_Cd^n\eta'|\eta - \eta'|^{3(2-n)} = 
\sum_{N=0}^{\infty}\sum_{\alpha = 0}^N
\frac{(2a)^{6-2n}{\pi}^{\frac{n}{2}}\Gamma(3-n)\Gamma(N-3+\frac{3n}{2})}{
\Gamma(N+3-\frac{n}{2})\Gamma(\frac{3n}{2}-3)}\sqrt{{a^n \Omega_{n+1}}}
Y_{a}^N({\eta})\times
$$

$$
\big{[}
-aK\frac{(sin{\theta_n^0})^{m_1}C_{N-m_1}^{m_1+\frac{n-1}{2}}(cos{\theta_n^0})
C_{0}^{\frac{n-1}{2}}(cos{\theta_n^0})}{(\frac{n}{2}-1)\frac{n}{2} - (N+\frac{n}{2}-1)(N+\frac{n}{2})}\delta^{m_10}\delta_{\alpha 0}\big{]}
$$
This expression is seen to contain a pole at the dimensional limit stemming from $ \Gamma(3-n)$. However, the particularity of the diagram in question is already evident in the presence of $ Y_{\alpha}^N(\eta)$ on the left-hand side of this expresion. It will become obvious that, unlike the situation with all other diagrams in fig.(1), such a space-time dependence can not be integrated out in a manner which preserves the structure of the connected diagram (1j) on $ C_4$. Resolving the stated spherical harmonic through (A8) and using (14) as well as  \cite{GT}

\begin{equation}
\delta_{\alpha\alpha'} = \delta^{m_1l_1}\delta_{m_2l_2}...\delta_{m_{n-1}l_{n-1}} =  
\delta^{m_1l_1}\delta_{aa'}
\end{equation}
before taking the limit $ n \rightarrow 4$ results in 

$$
\left[ \frac{\Gamma(\frac{n}{2} - 1)}{4\pi^{\frac{n}{2}}}\right]^3
\int_Cd^n\eta'|\eta - \eta'|^{3(2-n)} = 
$$

\begin{equation}
\frac{1}{2^9}\frac{1}{\pi^4}\big{[}\Omega_5\big{]}^{\frac{1}{2}}\sum_{N=1}^{\infty}\frac{(N + 1)(N + 2)}{N(N + 3)}C_N^{\frac{3}{2}}(cos{\theta_4^0})\big{[}a^{-2}C_N^{\frac{3}{2}}(cos{\theta_4})\big{]}
(a^{-2})a^3K \frac{1}{\epsilon} + F. T.
\end{equation}
with the powers of the embedding radius $ a$ having been conveniently distributed for the purposes of the ensuing calculation.   

The product of the first term in (27) with the second term in (26) yields

$$
\int_Cd^n\eta'|\eta - \eta'|^{3(2-n)}\frac{(\eta - \eta')^2}{2na^2}\frac{L^2}{2} =
$$

\begin{equation}
\frac{L^2}{4na^2}\sum_{N=0}^{\infty}\sum_{\alpha = 0}^N
\frac{(2a)^{8-2n}{\pi}^{\frac{n}{2}}\Gamma(4-n)\Gamma(N-4+\frac{3n}{2})}{
\Gamma(N+4-\frac{n}{2})\Gamma(\frac{3n}{2}-4)}Y_{\alpha}^N({\eta})\int_Cd^n\eta'Y_{\alpha}^N(\eta')
\end{equation}
which, on the same lines as those leading to (30), at $ n \rightarrow 4$ 
amounts to

$$
\left[ \frac{\Gamma(\frac{n}{2} - 1)}{4\pi^{\frac{n}{2}}}\right]^3
\int_Cd^n\eta'|\eta - \eta'|^{3(2-n)}\frac{(\eta - \eta')^2}{2na^2}\frac{L^2}{2} =
$$

\begin{equation}
-\frac{1}{2^{10}}\frac{1}{\pi^4}\big{[}\Omega_5\big{]}^{\frac{1}{2}}\sum_{N=1}^{\infty}\frac{1}{N(N + 3)}C_N^{\frac{3}{2}}(cos{\theta_4^0})\big{[}a^{-2}C_N^{\frac{3}{2}}(cos{\theta_4})\big{]}
(a^{-2}L^2)a^3K \frac{1}{\epsilon} + F. T.
\end{equation}

The additive result of (30) and (32) at $ n \rightarrow 4$ amounts to 

$$
\left[ \frac{\Gamma(\frac{n}{2} - 1)}{4\pi^{\frac{n}{2}}}\right]^3
\int_Cd^n\eta'\big{[}|\eta - \eta'|^{3(2-n)}\big{]}\big{[}1 + \frac{(\eta - \eta')^2}{2na^2}\frac{L^2}{2}\big{]} = 
$$

\begin{equation}
-\frac{1}{3}\frac{1}{2^{9}}\frac{1}{\pi^4}\big{[}\Omega_5\big{]}^{\frac{1}{2}}cot{\theta_4^0}\sum_{N=1}^{\infty}\frac{C_N^{\frac{3}{2}}(cos{\theta_4^0})}{N(N + 3)}C_N^{\frac{3}{2}}(cos{\theta_4})\big{[}\frac{L^2 - 2(N + 1)(N + 2)}{2a^{2}}\big{]}
\frac{1}{\epsilon} + F. T.
\end{equation}

The next term in order results, through (A1) and (A2), in 

$$
-3\int_Cd^n\eta'\big{[}|\eta - \eta'|^{2(2-n)}|\frac{a_{\eta'}}{a_B}\eta - \frac{a_B}{a_{\eta'}}\eta'|^{2-n}\big{]} = 
$$

$$
-3\sum_{N=0}^{\infty}\sum_{\alpha = 0}^N
\frac{(2a)^{4-n}{\pi}^{\frac{n}{2}}\Gamma(2-\frac{n}{2})\Gamma(N-2+n)}{
\Gamma(N+2)\Gamma(n-2)}\sum_{N'=0}^{N'_0}\sum_{\alpha'=0}^{N'}
\frac{(2a)^{2}{\pi}^{\frac{n}{2}}\Gamma(1+\frac{1}{N'_0})\Gamma(N'-1+\frac{n}{2}+\frac{1}{N'_0})}{\Gamma(N'+1+\frac{n}{2}+\frac{1}{N'_0})\Gamma(\frac{n}{2}-1)}\times
$$

\begin{equation}
Y_{\alpha}^N({\eta})Y_{\alpha'}^{N'}({\eta})\int_Cd^n\eta'\big{[}Y_{\alpha}^N(\eta')
Y_{\alpha'}^{N'}({\eta}')\big{]}
\end{equation}
As was the case in (28), again this expression manifests a specific space-time dependence
which will be shown to be incompatible with the mathematical structure of diagram (1j) on
$ C_4$. In the context of (A10') and (A11) the integral over $ \eta'$ in this expression is

$$
\int_Cd^n\eta'Y_{\alpha}^N(\eta')Y_{\alpha'}^{N'}({\eta}') = 
\frac{aKF(n)\delta_{\alpha\alpha'} + B(n)H(n)}{(N' + \frac{n}{2} - 1)(N' + \frac{n}{2}) - (N + \frac{n}{2} - 1)(N + \frac{n}{2})}
$$ 
whereby use of the relation between the extrinsic curvature $ K$ and the boundary-defining angle $ \theta_4^0$ \cite{G}

\begin{equation}
K = \frac{1}{3}\frac{1}{a}cot{\theta_4^0}
\end{equation}
results, at the limit $ n \rightarrow 4$, in 

$$
-3\left[ \frac{\Gamma(\frac{n}{2} - 1)}{4\pi^{\frac{n}{2}}}\right]^3
\int_Cd^n\eta'\big{[}|\eta - \eta'|^{2(2-n)}|\frac{a_{\eta'}}{a_B}\eta - \frac{a_B}{a_{\eta'}}\eta'|^{2-n}\big{]} =
-\frac{3}{\pi^2}\sum_{N=0}^{\infty}\sum_{N'=0}^{N'_0}
\frac{\Gamma(1+\frac{1}{N'_0})\Gamma(N'+1+\frac{1}{N'_0})}{\Gamma(N'+3+\frac{1}{N'_0})}
\times
$$

\begin{equation}
\frac{a^2}{N'(N'+3)-N(N+3)}\big{[}\frac{1}{3}cot{\theta_4^0}\sum_{\alpha =0}^{N'}F(4)Y_{\alpha}^{N}(\eta)Y_{\alpha}^{N'}(\eta)+[B(4)H(4)]\sum_{\alpha =0}^{N}\sum_{\alpha' =0}^{N'}Y_{\alpha}^{N}(\eta)Y_{\alpha'}^{N'}(\eta)\big{]}\frac{1}{\epsilon} + F. T.
\end{equation}
It is a trivial matter to confirm that all other terms in (26) are finite at $ n \rightarrow 4$. 

The local dependence of the residues in the expression for diagram (1j) on the Riemannian manifold $ C_4$ is manifest in (33) and (36). This is an unusual situation which necessitates caution. In order to arrive at a consistent physical interpretation a Taylor expansion will be considered about any fixed three-dimensional hypersurface of $ C_4$ characterised by an embedding azimouthal angle $ \theta_4^*$ in the expression of diagram (1j) which essentially amounts to the additive result of (33) and (36). With a symmetry factor of $ \frac{1}{6}$ the Taylor expansion in (33) yields

$$
\left[ \frac{\Gamma(\frac{n}{2} - 1)}{4\pi^{\frac{n}{2}}}\right]^3
\int_Cd^n\eta'\big{[}|\eta - \eta'|^{3(2-n)}\big{]}\big{[}1 + \frac{(\eta - \eta')^2}{2na^2}\frac{L^2}{2}\big{]} = $$

$$ 
-\frac{1}{3^2}\frac{1}{2^{10}}\frac{1}{\pi^4}\big{[}\Omega_5\big{]}^{\frac{1}{2}}cot{\theta_4^0}\sum_{N=1}^{\infty}\frac{C_N^{\frac{3}{2}}(cos{\theta_4^0})}{N(N + 3)}
\big{[}C_N^{\frac{3}{2}}(cos{\theta_4^*}) +
2\frac{\Gamma(\frac{5}{2})}{\Gamma(\frac{3}{2})}C_{N-1}^{\frac{5}{2}}(cos{\theta_4^*})
\big{[}cos\theta_4 - cos\theta_4^*\big{]} +
$$

\begin{equation}
4\frac{\Gamma(\frac{7}{2})}{\Gamma(\frac{5}{2})}C_{N-2}^{\frac{7}{2}}(cos{\theta_4^*})
\big{[}cos\theta_4 - cos\theta_4^*\big{]}^2 + ...\big{]}
\big{[}\frac{L^2 - 2(N + 1)(N + 2)}{2a^{2}}\big{]}
\frac{1}{\epsilon} + F. T.
\end{equation}
where use has been made of \cite{Bateman}

$$
\frac{d^k}{dt^k}C_n^{\lambda}(t) = 2^k\frac{\Gamma(\lambda+k)}{\Gamma(\lambda)}
C_{n-k}^{\lambda+k}(t)
$$
Use of (2) on the right-hand side of (37) can be readily seen to result in

$$
\left[ \frac{\Gamma(\frac{n}{2} - 1)}{4\pi^{\frac{n}{2}}}\right]^3
\int_Cd^n\eta'\big{[}|\eta - \eta'|^{3(2-n)}\big{]}\big{[}1 + \frac{(\eta - \eta')^2}{2na^2}\frac{L^2}{2}\big{]} = 
$$

$$
-\frac{1}{3^2}\frac{1}{2^{10}}\frac{1}{\pi^4}\big{[}\Omega_5\big{]}^{\frac{1}{2}}cot{\theta_4^0}\sum_{N=1}^{\infty}\frac{C_N^{\frac{3}{2}}(cos{\theta_4^0})}{N(N + 3)}
C_N^{\frac{3}{2}}(cos{\theta_4^*})\big{[}\frac{L^2 - 4}{2a^{2}}\big{]}\frac{1}{\epsilon} 
$$

$$
-\frac{1}{3^2}\frac{1}{2^{10}}\frac{1}{\pi^4}\big{[}\Omega_5\big{]}^{\frac{1}{2}}cot{\theta_4^0}\sum_{N=1}^{\infty}\frac{C_N^{\frac{3}{2}}(cos{\theta_4^0})}{N(N + 3)}\big{[}
3C_{N-1}^{\frac{5}{2}}(cos{\theta_4^*})(cos\theta_4 - cos\theta_4^*) +
$$

$$
10C_{N-2}^{\frac{7}{2}}(cos{\theta_4^*})(cos\theta_4 - cos\theta_4^*)^2 + ...\big{]}
\big{[}\frac{L^2 - 4}{2a^{2}}\big{]}
\frac{1}{\epsilon} +  
$$

$$
\frac{1}{3^3}\frac{1}{2^{12}}\frac{1}{\pi^4}\big{[}\Omega_5\big{]}^{\frac{1}{2}}cot{\theta_4^0}\sum_{N=1}^{\infty}C_N^{\frac{3}{2}}(cos{\theta_4^0})C_N^{\frac{3}{2}}(cos{\theta_4^*})
R\frac{1}{\epsilon} +
$$

$$
\frac{1}{3^3}\frac{1}{2^{12}}\frac{1}{\pi^4}\big{[}\Omega_5\big{]}^{\frac{1}{2}}cot{\theta_4^0}\sum_{N=1}^{\infty}C_N^{\frac{3}{2}}(cos{\theta_4^0})\big{[}
3C_{N-1}^{\frac{5}{2}}(cos{\theta_4^*})(cos\theta_4 - cos\theta_4^*) +
$$

\begin{equation}
10C_{N-2}^{\frac{7}{2}}(cos{\theta_4^*})(cos\theta_4 - cos\theta_4^*)^2 + ...\big{]}R
\frac{1}{\epsilon} 
\end{equation}

The contribution to the two-point function generated by diagram (1j) is, in effect, the additive result of (38) and of the expression arising through the Taylor expansion about $ \theta_4^*$ in (36) after replacing $ a^2$ by the scalar curvature $ R$ through (2). Unlike the Taylor expansion in (38) however, clearly the residues stemming from each individual term in the expansion related to (36), including the zero-order term, depend locally, through $ \theta_i, i=1,2,3$, on the hypersurface of $ C_4$ specified by $ \theta_4^*$.

{\bf III. Renormalisation and gravitational backreaction.}

In the context of the situation hitherto described it is evident that any term in (38) which features a dependence on $ cos{\theta_4}$ signifies a radiative contribution to the scalar two-point function on a background which deviates from the geometry of $ C_4$.
Specifically, substituting (38) in (26) and comparing with (3) reveals that terms of the form 

\begin{equation}
\frac{1}{\epsilon}
\int_Cd^n\eta D_c(\eta_1 - \eta)\big{[}\frac{L^2 - \frac{1}{2}n(n-2)}{2a^2}\big{]}_{n=4}
(cos{\theta_4} - cos{\theta_4^*})^lD_c(\eta - \eta_2);~~~l \in N 
\end{equation}
as well as of the form

\begin{equation}
\frac{1}{\epsilon}
\int_Cd^n\eta D_c(\eta_1 - \eta)\big{[}R(cos{\theta_4} - cos{\theta_4^*})^{l'}\big{]}_{n=4}
D_c(\eta - \eta_2);~~~l' \in N 
\end{equation}
signify, order by order in $ l$ and $ l'$, a deviation from the kernel   

$$ 
\frac{L^2 - \frac{1}{2}n(n-2)}{2a^2}
$$
and from the positive constant scalar curvature $ R$. Such deviations do not explicitly specify the modified geometry on which the scalar vacuum effects are realised since the local space-time dependence inherent in any power of the difference in the cosines is essentially the result of a combination between, for instance, the resulting local curvature $ R(x)$ and the resulting $ [-g(x)]^{\frac{1}{2}}$ inherent in the invariant integration measures. The terms in (39) and (40) do, however, signify deviations from the background geometry of $ C_4$ the degree of which diminishes with any increase of the powers $ l$ and $ l'$. For the same reason, each term in the Taylor expansion associated with (36) signifies individually a geometric background which detracts from that of $ C_4$ since, as stated, all associated residues depend locally on the three-dimensional hypersurface of $ C_4$ specified by $ \theta_4$. Consequently, the only two terms in (38) signifying radiative contributions to the two-point function on $ C_4$ associated with diagram (1j) are seen through (26) to result in

$$ 
(1j):
$$

$$
-\lambda^2\frac{1}{3^2}\frac{1}{2^{10}}\frac{1}{\pi^4}\big{[}\Omega_5\big{]}^{\frac{1}{2}}cot{\theta_4^0}\sum_{N=1}^{\infty}\frac{C_N^{\frac{3}{2}}(cos{\theta_4^0})}{N(N + 3)}
C_N^{\frac{3}{2}}(cos{\theta_4})\frac{1}{\epsilon}\int_Cd^4\eta D_c(\eta_1 - \eta)
\big{[}\frac{L^2 - 4}{2a^{2}}\big{]}D_c(\eta - \eta_2) 
$$  

\begin{equation}
\hspace{0.5in}
+\lambda\frac{1}{3^3}\frac{1}{2^{12}}\frac{1}{\pi^4}\big{[}\Omega_5\big{]}^{\frac{1}{2}}cot{\theta_4^0}\sum_{N=1}^{\infty}C_N^{\frac{3}{2}}(cos{\theta_4^0})C_N^{\frac{3}{2}}(cos{\theta_4})R\frac{1}{\epsilon}\int_Cd^4\eta D_c(\eta_1 - \eta)D_c(\eta - \eta_2)
\end{equation}
where the label $ *$ specifying any hypersurface on $ C_4$ has been dropped. This is the desired contribution of diagram (1j) on $ C_4$ which completes the list citing the expressions (21)-(25) and (A5) for the diagrams relevant to the renormalisation procedure to second order in the scalar self-coupling. 

In addition to the exceptional feature of the partial contribution which (41) signifies for (1j) on $ C_4$ an additional unusual trait for the associated radiative contribution is immediately obvious in that expression. Specifically, the residues manifest a dependence on $ \theta_4$ and they are, for that matter, functions of imaginary time on the Riemannian $ C_4$, assuming an equal time foliation of the manifold. Although in general expected this situation detracts from the context of matter field renormalisation on a fixed semi-classical geometrical background of constant curvature in which all residues have no dependence on the background space-time \cite{Drummond}, 
\cite{McKeon Tsoupros}, \cite{OBS}. The absence of such a dependence can be intuitively understood through the nature of the associated divergence as an effect arising from the coincidence of points on the background manifold. Such an effect should not be affected by the background constant space-time curvature. In general, however, a divergence is known to receive contributions, locally, from the background curvature \cite{Birrel}, \cite{OBS}
and, globally, from any boundary present \cite{GT}, \cite{AvraEspo}, \cite{Toms}.
For that matter, in the context of a general manifold the absence of the stated dependence in the residues is expected to persist at least as a leading divergence. Notwithstanding such an expectation, however, the local dependence of the residues as well as the stated partial relevance of diagram (1j) on $ C_4$ reveal the effects of the gravitational backreaction generated by the scalar vacuum effects on the geometry of that manifold. Specifically, the essence of the semi-classical approach to Quantum Gravity in which all matter fields are quantised on a fixed semi-classical space-time geometry is expressed by replacing the stress tensor in Einstein's field equations by its vacuum expectation value \cite{Birrel}

\begin{equation}        
G_{\mu\nu} = -8\pi G_0<0_{out}|T_{\mu\nu}|0_{in}>
\end{equation}
The the right-hand side of (42) although always signifying a classical geometry does not, however, necessarily remain what it was prior to the replacement of the stress tensor by its expectation value. Quantum fluctuations of the matter fields inherent in that vacuum expectation value have the potential to affect the initial geometry. Such backreaction effects correspond to perturbative contributions on the initial Einstein tensor of the classical theory. In the present case, the backreaction effects on the geometry of $ C_4$ occur at two-loop order through the scalar two-point function and are responsible for     
the stated deviations from the classical geometry manifest in (39) and (40). For the same reason they are responsible for the local dependence of the associated residues on imaginary time, in the context of an equal time foliation of $ C_4$. Again, this feature can be intuitively understood by the fact that, as stated, in general the background curvature will affect the divergences to result in  space-time dependend residues \cite{Birrel} which allow for a general, space-time independent, leading divergence. Such is the present case at length scales relevant to two-loop scalar vacuum effects. The background geometry at such length scales ceases to be that of the positive constant curvature of $ C_4$ due to the gravitational backreaction. However, backreaction effects continue to affect the dynamical behaviour of the scalar field on $ C_4$ at larger scales generating, at two loop-order, the terms  featured in (41) whose dependence on  imaginary time is indicative of the deviations from the geometry of constant curvature. For this reason the results obtained herein at two-loop order differ, in this respect, from the results obtained through the evaluation of the one-loop effective action \cite{Od}, \cite{intsov}.    

The contribution of radiative effects to the bare self-coupling has been evaluated to second order in (20). Having established, to that order, the results for all relevant diagrams use will be made of the expansions in (6) and (7) in order to determine the contributions to the remaining bare parameters. To that effect, the relevant contribution $ I^{(2)}(\eta_1, \eta_2)$ to the renormalised two-point function $ G_R^{(2)}(\eta_1, \eta_2)$ is perturbatively expressed by the sum-total of the mathematical expressions of diagrams (1d), (1i) and (1j). As the associated expressions are substantially involved the calculation resulting in the  renormalisation of the two-point function is given in some detail 

\begin{equation}
I^{(2)}(\eta_1, \eta_2) = 
\end{equation}

$$
D_c(\eta_1 - \eta_2) + \lambda^2\big{[}\frac{1}{2^{11}}\frac{1}{\pi^\frac{11}{2}}(V_c + 1)\sum_{N=1}^{\infty}\frac{[C_N^{\frac{3}{2}}(cos{\theta_4^0})]^2}{N^2(N + 3)^2}\sum_{N'=0}^{N'_0}(2N' + 3)[\frac{3}{2}(V_c + 1) - 1]K^2 $$  

$$
+ \frac{1}{3^3}\frac{1}{2^{12}}\frac{1}{\pi^4}\big{[}\Omega_5\big{]}^{\frac{1}{2}}cot{\theta_4^0}\sum_{N=1}^{\infty}C_N^{\frac{3}{2}}(cos{\theta_4^0})C_N^{\frac{3}{2}}(cos{\theta_4})R\big{]}\frac{1}{\epsilon}\int_Cd^4\eta D_c(\eta_1 - \eta)D_c(\eta - \eta_2) 
$$

$$
-\lambda^2 \frac{1}{3^2}\frac{1}{2^{10}}\frac{1}{\pi^4}\big{[}\Omega_5\big{]}^{\frac{1}{2}}cot{\theta_4^0}\sum_{N=1}^{\infty}\frac{C_N^{\frac{3}{2}}(cos{\theta_4^0})}{N(N + 3)}
C_N^{\frac{3}{2}}(cos{\theta_4})\frac{1}{\epsilon}\int_Cd^4\eta D_c(\eta_1 - \eta)
\big{[}\frac{L^2 -\frac{1}{2}n(n - 2)}{2a^{2}}\big{]}_{n \rightarrow 4}D_c(\eta - \eta_2)  
$$

$$
+ F. T.
$$
Observing that to $ O(\lambda^2)$ it is 

$$
\big{[}1 - \lambda^2 \frac{1}{3^2}\frac{1}{2^{10}}\frac{1}{\pi^4}\big{[}\Omega_5\big{]}^{\frac{1}{2}}cot{\theta_4^0}\sum_{N=1}^{\infty}\frac{C_N^{\frac{3}{2}}(cos{\theta_4^0})}{N(N + 3)}
C_N^{\frac{3}{2}}(cos{\theta_4})\frac{1}{\epsilon}\big{]}\times
$$

$$
\big{[}1 + \lambda^2 
\frac{1}{3^2}\frac{1}{2^{10}}\frac{1}{\pi^4}\big{[}\Omega_5\big{]}^{\frac{1}{2}}cot{\theta_4^0}\sum_{N=1}^{\infty}\frac{C_N^{\frac{3}{2}}(cos{\theta_4^0})}{N(N + 3)}
C_N^{\frac{3}{2}}(cos{\theta_4})\frac{1}{\epsilon}\big{]} = 1
$$
allows for that expression in the form

$$
I^{(2)}(\eta_1, \eta_2) = 
\big{(}1 - \lambda^2 \frac{1}{3^2}\frac{1}{2^{10}}\frac{1}{\pi^4}\big{[}\Omega_5\big{]}^{\frac{1}{2}}cot{\theta_4^0}\sum_{N=1}^{\infty}\frac{C_N^{\frac{3}{2}}(cos{\theta_4^0})}{N(N + 3)}
C_N^{\frac{3}{2}}(cos{\theta_4})\frac{1}{\epsilon}\big{)}\times
$$

$$
\lambda^2\big{[}\big{(}\frac{1}{2^{11}}\frac{1}{\pi^\frac{11}{2}}(V_c + 1)\sum_{N=1}^{\infty}\frac{[C_N^{\frac{3}{2}}(cos{\theta_4^0})]^2}{N^2(N + 3)^2}\sum_{N'=0}^{N'_0}(2N' + 3)[\frac{3}{2}(V_c + 1) - 1]\big{)}K^2\frac{1}{\epsilon}\times $$  

$$
\big{(}1 + \lambda^2 \frac{1}{3^2}\frac{1}{2^{10}}\frac{1}{\pi^4}\big{[}\Omega_5\big{]}^{\frac{1}{2}}cot{\theta_4^0}\sum_{N=1}^{\infty}\frac{C_N^{\frac{3}{2}}(cos{\theta_4^0})}{N(N + 3)}
C_N^{\frac{3}{2}}(cos{\theta_4})\frac{1}{\epsilon}\big{)}
$$

$$
+ \big{(}\frac{1}{3^3}\frac{1}{2^{12}}\frac{1}{\pi^4}\big{[}\Omega_5\big{]}^{\frac{1}{2}}cot{\theta_4^0}\sum_{N=1}^{\infty}C_N^{\frac{3}{2}}(cos{\theta_4^0})C_N^{\frac{3}{2}}(cos{\theta_4})\big{)}R\frac{1}{\epsilon}\times
$$

\begin{equation}
\big{(}1 + \lambda^2 \frac{1}{3^2}\frac{1}{2^{10}}\frac{1}{\pi^4}\big{[}\Omega_5\big{]}^{\frac{1}{2}}cot{\theta_4^0}\sum_{N=1}^{\infty}\frac{C_N^{\frac{3}{2}}(cos{\theta_4^0})}{N(N + 3)}
C_N^{\frac{3}{2}}(cos{\theta_4})\frac{1}{\epsilon}\big{)} + 1\big{]}D_c(\eta_1 - \eta_2) 
\end{equation}
where use has been made of 

$$
\int_Cd^4\eta D_c(\eta_1 - \eta)D_c(\eta - \eta_2) = D_c(\eta_1 - \eta_2)
$$
and 

$$
\int_Cd^4\eta D_c(\eta_1 - \eta)\big{[}\frac{L^2 - 4}{2a^{2}}\big{]}D_c(\eta - \eta_2) = 
\int_Cd^4\eta D_c(\eta_1 - \eta)\delta^{(4)}(\eta - \eta_2) = 
D_c(\eta_1 - \eta_2)
$$
The renormalised two-point function $ G_R^{(2)}(\eta_1 - \eta_2)$ emerges from (44) through the addition of the relevant counterterms. The determination of the residues relevant to the bare parameters associated with the two-point function are determined by the demand that the addition of appropriate counterterms in each sector yield a finite result. Specifically, for the $ K^2$-sector of (44) - which emerged as the sum-total of diagrams (1d) and (1i) - such a demand, in the context of (6d), amounts to

$$
(1d) + (1i) + (-)\frac{d_{12}\lambda^2}{\epsilon} < \infty
$$
and, likewise, the addition of the counterterm associated with (6c) in the $ R$-sector of 
(44) should yield a finite result. The remaining divergence should be absorbed in a redefinition of the scalar field realised in the context of (6f). Equivalently, the demand that 
the renormalised two-point function  
be finite to $ O(\lambda^2)$ translates to the demand that 

$$    
Z^{-1}\big{(}1 - \lambda^2 \frac{1}{3^2}\frac{1}{2^{10}}\frac{1}{\pi^4}\big{[}\Omega_5\big{]}^{\frac{1}{2}}cot{\theta_4^0}\sum_{N=1}^{\infty}\frac{C_N^{\frac{3}{2}}(cos{\theta_4^0})}{N(N + 3)}
C_N^{\frac{3}{2}}(cos{\theta_4})\frac{1}{\epsilon}\big{)}\times
$$

$$
\lambda^2\big{[}\big{(}\frac{1}{2^{11}}\frac{1}{\pi^\frac{11}{2}}(V_c + 1)\sum_{N=1}^{\infty}\frac{[C_N^{\frac{3}{2}}(cos{\theta_4^0})]^2}{N^2(N + 3)^2}\sum_{N'=0}^{N'_0}(2N' + 3)[\frac{3}{2}(V_c + 1) - 1] - d_{12}\big{)}K^2\frac{1}{\epsilon}\times $$  

$$
\big{(}1 + \lambda^2 \frac{1}{3^2}\frac{1}{2^{10}}\frac{1}{\pi^4}\big{[}\Omega_5\big{]}^{\frac{1}{2}}cot{\theta_4^0}\sum_{N=1}^{\infty}\frac{C_N^{\frac{3}{2}}(cos{\theta_4^0})}{N(N + 3)}
C_N^{\frac{3}{2}}(cos{\theta_4})\frac{1}{\epsilon}\big{)}
$$

$$
+ \big{(}\frac{1}{3^3}\frac{1}{2^{12}}\frac{1}{\pi^4}\big{[}\Omega_5\big{]}^{\frac{1}{2}}cot{\theta_4^0}\sum_{N=1}^{\infty}C_N^{\frac{3}{2}}(cos{\theta_4^0})C_N^{\frac{3}{2}}(cos{\theta_4}) - c_{12}\big{)}R\frac{1}{\epsilon}\times
$$

\begin{equation}
\big{(}1 + \lambda^2 \frac{1}{3^2}\frac{1}{2^{10}}\frac{1}{\pi^4}\big{[}\Omega_5\big{]}^{\frac{1}{2}}cot{\theta_4^0}\sum_{N=1}^{\infty}\frac{C_N^{\frac{3}{2}}(cos{\theta_4^0})}{N(N + 3)}
C_N^{\frac{3}{2}}(cos{\theta_4})\frac{1}{\epsilon}\big{)} + 1\big{]}D_c(\eta_1 - \eta_2) 
\end{equation}
be finite at $ n\rightarrow 4$. That, in effect, is equivalent to demanding that

\begin{equation}
d_{12}\lambda^2 
= \lambda^2\frac{1}{2^{11}}\frac{1}{\pi^\frac{11}{2}}(V_c + 1)\sum_{N=1}^{\infty}\frac{[C_N^{\frac{3}{2}}(cos{\theta_4^0})]^2}{N^2(N + 3)^2}\sum_{N'=0}^{N'_0}(2N' + 3)[\frac{3}{2}(V_c + 1) - 1]
\end{equation}

\begin{equation}
c_{12}\lambda^2 = \lambda^2\frac{1}{3^3}\frac{1}{2^{12}}\frac{1}{\pi^4}\big{[}\Omega_5\big{]}^{\frac{1}{2}}cot{\theta_4^0}\sum_{N=1}^{\infty}C_N^{\frac{3}{2}}(cos{\theta_4^0})C_N^{\frac{3}{2}}(cos{\theta_4})
\end{equation}
as well as, to $ O(\lambda^2)$ 

$$
Z = 
1 - \lambda^2 \frac{1}{3^2}\frac{1}{2^{10}}\frac{1}{\pi^4}\big{[}\Omega_5\big{]}^{\frac{1}{2}}cot{\theta_4^0}\sum_{N=1}^{\infty}\frac{C_N^{\frac{3}{2}}(cos{\theta_4^0})}{N(N + 3)}
C_N^{\frac{3}{2}}(cos{\theta_4})\frac{1}{\epsilon}
$$

Finally, the renormalisation in the gravitational sector of the bare action relates to the vacuum contributions associated with diagrams (1g), (1e) and (1f) which to $ O(\lambda^2)$ are, respectively, expressed by (A5), (22) and (23). The sum-total of them amounts to

$$
\lambda^2
\frac{1}{\pi^{4}}\frac{1}{3}\frac{1}{2^{12}}\sum_{N=0}^{\infty}\big{(}
\frac{1}{\pi^{2}}\frac{1}{3^3}\frac{1}{2^7}\frac{(N+2)(N+3)}{(N+1)(N+4)} 
\big{(}C_{N+1}^{\frac{3}{2}}(cos{\theta_4^0})\big{)}^2 + $$

$$
\sum_{N'=0}^{N'_0}\frac{1}{[{N'}^2-N^2+3(N'-N)]^2}[\Gamma(\frac{1}{N'_0}) +
\frac{1}{3}\frac{1}{2}\frac{\Gamma(1+\frac{1}{N'_0})\Gamma(N'+1+\frac{1}{N'_0})}{\Gamma(N'+3+\frac{1}{N'_0})}(N+1)(N+2)]\times $$

$$
\big{[}\frac{3}{2^3}\frac{1}{\pi^2}
F^2\big{]} - 
\big{[}\frac{3}{2^{9}}\frac{(V_c + 1)^2}{\pi^5} + \frac{\pi^3}{2^{5}}(V_c + 1)\big{]}
\big{[}\sum_{N'=0}^{N'_0}(2N' + 3)\big{]}^2
\sum_{N=1}^{\infty}\frac{[C_N^{\frac{3}{2}}(cos{\theta_4^0})]^2}{N^2(N+3)^2}\big{)} 
RK^2\frac{1}{\epsilon}\int_Cd^n\eta + 
$$

$$
\lambda^2\frac{1}{\pi^{6}}\frac{1}{3}\frac{1}{2^{17}}\sum_{N=0}^{\infty}\sum_{N'=0}^{N'_0}
\frac{1}{[{N'}^2-N^2+3(N'-N)]^2} \times $$

$$
\left[ \Gamma(\frac{1}{N'_0}) + 
\frac{1}{3}\frac{1}{2}\frac{\Gamma(1+\frac{1}{N'_0})\Gamma(N'+1+\frac{1}{N'_0})}{\Gamma(N'+3+\frac{1}{N'_0})}(N+1)(N+2) \right](BH)^2
\frac{R^2}{\epsilon}\int_Cd^4{\eta} +  $$

$$
\lambda^2\frac{1}{\pi^{6}}\frac{1}{3}\frac{1}{2^{12}}\sum_{N=0}^{\infty}\sum_{N'=0}^{N'_0}
\frac{1}{[{N'}^2-N^2+3(N'-N)]^2} \times $$

$$
\left[ \Gamma(\frac{1}{N'_0}) + 
\frac{1}{3}\frac{1}{2}\frac{\Gamma(1+\frac{1}{N'_0})\Gamma(N'+1+\frac{1}{N'_0})}{\Gamma(N'+3+\frac{1}{N'_0})}(N+1)(N+2) \right]\times $$

\begin{equation}
\hspace{1.5in}
(FBH)\big{(}sin(\theta_4^0)\big{)}^{-3}
\frac{RK}{\epsilon}\oint_{\partial C}d^3{\eta} 
\hspace{2.0in}
\end{equation}
It is obvious that the multiplicative factor featured in the residue of the $ RK^2\int_cd^4\eta$ sector in (48) yields an infinite contribution to same sector of the bare gravitational action in(4). Specifically, it yields the $ \beta_{12}\lambda^2$ contribution in the expansion (7b) of the bare gravitational coupling $ \beta_0$. The $ O(\lambda^2)$ contributions to the gravitational couplings $ \alpha_0$ and $ \gamma_0$ in expansions (7a) and (7c) can be, likewise, read off the $ R^2\int_cd^4\eta$ and $ RK\oint_cd^3\eta$ sectors of (48) respectively.

{\bf IV. Conclusion}

The renormalisation of a second quantised conformally invariant scalar field theory on $ C_4$, the Riemannian four-dimensional  manifold of positive constant curvature characterised by a boundary of constant positive extrinsic curvature, has been considered to second order in the scalar self-coupling in the context of the Dirichlet condition of a vanishing scalar field on the boundary. The method of images has been the basis of all diagramatic calculations. The association of the dynamical behaviour of the stated field to that on the Euclidean de Sitter space $ S_4$, for which that method allows, effectively results in a closed expression for the scalar propagator which, in a general space-time, receives contributions from actual gravitons to all orders in the metric tensor. In the context of dimensional regularisation there are no infinite contributions to the Einstein-Hilbert action at one-loop level. To second order in the scalar self-coupling scalar vacuum effects result in divergences which necessitate the introduction of $ RK^2$ and $ R^2$ volume terms as well as an $ RK$ surface term in the gravitational component of the effective action. At two-loop level, substantial backreaction effects occur through radiative effects in the two-point function. In modifying the background geometry such effects signal the scale at which scalar loop effects begin to dissociate themselves from the renormalisation procedure on $ C_4$. However, the modified geometry contributes to the radiative effects of the two-point function on $ C_4$ at larger scales. Although at one and two-loop order no mass generation emerges as a result of such effects to second order in the scalar self-coupling radiative effects induce a wave function renormalisation for the scalar field as well as terms proportional to $ R\Phi^2$ and $ K^2\Phi^2$ in the effective action. Such terms explicitly break conformal invariance at two-loop level.                  

{\bf Acknowledgements}

I would like to express my appreciation to the University of Beijing for its kind invitation. I would, also, like to thank Professor Liu Liao for an interesting conversation over my results. As science is humanity's highest endeavor, I would like 
to dedicate this work to the heroic resistance of the people of Iraq against the 
invaders of their country. The fighting humanity's sense of freedom makes science always worth pursuing. 

\newpage

{\bf Appendix}

The evaluation of the diagramatic structures relevant to the renormalisation procedure 
necessitates the results obtained through the spherical formulation for the mathematical 
expressions of Feynman diagrams on $ C_n$ \cite{T}, \cite{GT}. Of fundamental importance are 

$$
(A1)\hspace{1.0in}
[(\eta - \eta')^2]^{\nu} = \sum_{N=0}^{\infty}\sum_{\alpha=0}^{N}
\frac{(2a)^{2\nu+n}{\pi}^{\frac{n}{2}}\Gamma(\nu+\frac{n}{2})\Gamma(N-\nu)}{
\Gamma(N+n+\nu)\Gamma(-\nu)}Y_{\alpha}^N({\eta})Y_{\alpha}^N({\eta}')
\hspace{0.5in}
$$

$$
(A2)\hspace{0.1in}
[|\frac{a_{{\eta}'}}{a_B}{\eta}- \frac{a_B}{a_{{\eta}'}}{\eta}'|^2]^{\nu} = 
\sum_{N=0}^{N_0}\sum_{\alpha=0}^{N}
\frac{(2a)^{2\nu+n}{\pi}^{\frac{n}{2}}\Gamma(\nu+\frac{n}{2}+
\frac{1}{N_0})\Gamma(N-\nu+ \frac{1}{N_0})}{
\Gamma(N+n+\nu+ \frac{1}{N_0})\Gamma(-\nu)}Y_{\alpha}^N({\eta})Y_{\alpha}^N({\eta}')
$$
with the expressions on the left-hand sides of (A1) and (A2) stemming, respectively, from the singular and boundary part of the propagator $ D_c(\eta - \eta')$ which has been derived in \cite{G} and reproduced in expression (8) of the present project. In addition to the singular part which signifies the propagator on $ S_n$ the Green function in (8) also features a boundary part which admits itself the physical interpretation of the propagator between the image-points $ \frac{a_{{\eta}'}}{a_B}{\eta}$ and $
\frac{a_B}{a_{{\eta}'}}{\eta}'$. The presence of a lower non-vanishing limit for image propagation stemming from the boundary part at $ \eta \rightarrow \eta'$ enforces the associated ultra-violet cut-off $ N_0$ featured in (A2) \cite{G}.    

Allowing for a multiplicative factor the expression for diagram (1d) has been derived in \cite{G}. The singular part of the propagator $ D_c(\eta - \eta)$ is identical to the mathematical expression for the massless tadpole on $ S_n$ and, for that matter, it vanishes. In effect, in the context of dimensional regularisation, the loop structure in (1d) is described by the finite expression 

$$(A3)\hspace{1.0in}
\int_Cd^n\eta D_{c}^{(n)}(\eta,{\eta}) = \frac{\Gamma(\frac{n}{2}-1)}{4\pi^{\frac{n}{2}}}
\int_Cd^n\eta \big{[}- {|\frac{a_{{\eta}}}{a_B}{\eta}-
\frac{a_B}{a_{{\eta}}}{\eta}|^{2-n}}\big{]}
\hspace{2.0in}
$$
which stems from the boundary part of the propagator in (8). The exact expression for diagram (1d) including the symmetry factor of $ \frac{1}{2}$ and the bare self-coupling $ \lambda_0$ is 

$$
(A3')\hspace{0.5in}(1d):   
-(-\lambda_0)\frac{1}{3}\frac{1}{2^{10}}\frac{V_c + 1}{\pi^{\frac{7}{2}}}\big{[}\sum_{N=0}^{N_0}(2N + 
3)\big{]}R\int_Cd^4\eta D_c(\eta_1 - \eta)D_c(\eta - \eta_2)
\hspace{0.5in}
$$
with the volume of $ C_4$ relating to that of $ S_4$ as \cite{GT}

$$
(A4)\hspace{2.5in}
\int_{S_4}d^4\eta = (V_c + 1)\int_{C}d^4\eta~~~;~~~V_c > 0
\hspace{2.0in}
$$ 
and with the volume of the n-dimensional Euclidean de Sitter space of embedding radius $ a$  being 

$$
(A4')\hspace{2.5in}
\int_{S_n}d^n\eta = a^n\Omega_{n+1}
\hspace{2.0in}
$$

The perturbative contribution to the zero-point function stemming from diagram (1g) has been derived in \cite{GT}. With a boundary-specifying angle $ \theta_{4}^0$ and with
$ \epsilon \rightarrow 0$ as $ n \rightarrow 4$ its exact expression, including the associated symmetry factor of $ \frac{1}{48}$ and the bare self-coulping, is

$$
(1g):~~~
\lambda_0^2\frac{1}{\pi^{4}}\frac{1}{3}\frac{1}{2^{12}}\sum_{N=0}^{\infty}\big{[}
\frac{1}{\pi^{2}}\frac{1}{3^3}\frac{1}{2^7}\frac{(N+2)(N+3)}{(N+1)(N+4)} 
\big{(}C_{N+1}^{\frac{3}{2}}(cos{\theta_4^0})\big{)}^2 + $$

$$
\sum_{N'=0}^{N'_0}\frac{1}{[{N'}^2-N^2+3(N'-N)]^2}[\Gamma(\frac{1}{N'_0}) +
\frac{1}{3}\frac{1}{2}\frac{\Gamma(1+\frac{1}{N'_0})\Gamma(N'+1+\frac{1}{N'_0})}{\Gamma(N'+3+\frac{1}{N'_0})}(N+1)(N+2)] \times $$

$$
3\frac{1}{2^3}\frac{1}{\pi^2}
F^2\big{]}\frac{RK^2}{\epsilon}\int_Cd^4{\eta} + $$

$$
\lambda_0^2\frac{1}{\pi^{6}}\frac{1}{3}\frac{1}{2^{17}}\sum_{N=0}^{\infty}\sum_{N'=0}^{N'_0}
\frac{1}{[{N'}^2-N^2+3(N'-N)]^2} \times $$

$$
\left[ \Gamma(\frac{1}{N'_0}) + 
\frac{1}{3}\frac{1}{2}\frac{\Gamma(1+\frac{1}{N'_0})\Gamma(N'+1+\frac{1}{N'_0})}{\Gamma(N'+3+\frac{1}{N'_0})}(N+1)(N+2) \right](BH)^2
\frac{R^2}{\epsilon}\int_Cd^4{\eta} +  $$

$$
\lambda_0^2\frac{1}{\pi^{6}}\frac{1}{3}\frac{1}{2^{12}}\sum_{N=0}^{\infty}\sum_{N'=0}^{N'_0}
\frac{1}{[{N'}^2-N^2+3(N'-N)]^2} \times $$

$$
\left[ \Gamma(\frac{1}{N'_0}) + 
\frac{1}{3}\frac{1}{2}\frac{\Gamma(1+\frac{1}{N'_0})\Gamma(N'+1+\frac{1}{N'_0})}{\Gamma(N'+3+\frac{1}{N'_0})}(N+1)(N+2) \right]\times $$

$$
(A5)\hspace{2.0in}
(FBH)\big{(}sin(\theta_4^0)\big{)}^{-3}
\frac{RK}{\epsilon}\oint_{\partial C}d^3{\eta} 
\hspace{2.0in}
$$
with the condition $ N \neq N'$ and with $ n=4$ in all n-dependent quantities. The n-dependent quantities $ F, B, H$ which appear in (A5) are, at $ n \rightarrow 4$, respectively expressed by (A6), (A10) and (A12) which appear in what follows. 

$$
(A6)\hspace{1.0in}
F(4) = 
\sum_{m_1=0}^{N'cos{\theta_4^0}}(sin{\theta_4^0})^{2m_1}C_{N-m_1}^{m_1+\frac{3}{2}}(cos{\theta_4^0})
C_{N'-m_1}^{m_1+\frac{3}{2}}(cos{\theta_4^0})m_1
\hspace{1.0in}
$$
with $ \theta_n^0$ being the angle specifying $ \partial C_n$ in the $ (n+1)$-dimensional embedding Euclidean space and  

$$
(A7)\hspace{2.0in}
m_1^0 = Ncos{\theta_n^0}~~;~~l_1^0 = N'cos{\theta_n^0}
\hspace{2.0in}
$$
being the degrees of spherical harmonics defined on $ \partial C_n$. The Gegenbauer polynomials $ C_m^p(cos{\theta_n})$ with their explicit dependence on the angles associated with the embedding $ (n + 1)$-vector $ \eta$ relate to the n-dimensional spherical harmonics $ Y_{\alpha}^N(\eta)$ on $ S_n$ through

$$
(A8)\hspace{0.5in}Y_{\alpha}^N(\eta) = a^{-\frac{n}{2}}e^{\pm im_{n-1}\theta_1}
\prod_{k=0}^{n-2}(sin{\theta}_{n-k})^{m_{k+1}}C_{m_k-m_{k+1}}^{m_{k+1}+\frac{n-1}{2}-\frac{1}{2}k}(cos{\theta}_{n-k})
\hspace{1.0in}
$$
which, through the reduction formula  

$$
(A9)\hspace{0.5in}
\int_C d^{n}{\eta}Y_{\alpha}^N(\eta)Y_{\alpha'}^{N'}(\eta) = Aa^2
\oint_{\partial C} d^{n-1}{\eta}[KY_{\alpha}^N(\eta)Y_{\alpha'}^{N'}(\eta) + 
2n_pY_{\alpha'}^{N'}(\eta)D_pY_{\alpha}^{N}(\eta)]
$$
involving 

$$
D_a = (\delta_{ab} - \frac{\eta_a\eta_b}{a^2})\frac{\partial}{\partial \eta_a}
$$
and 

$$ 
K = \frac{1}{2}(D_in_j + D_jn_i)
$$
gives rise to

$$
\int_C d^{n}{\eta}Y_{\alpha}^N(\eta)Y_{\alpha'}^{N'}(\eta) = 
AKa
(sin{\theta_n^0})^{m_1+l_1}C_{N-m_1}^{m_1+\frac{n-1}{2}}(cos{\theta_n^0})
C_{N'-l_1}^{l_1+\frac{n-1}{2}}(cos{\theta_n^0})\delta^{m_1l_1}\delta_{aa'}~~ +  
$$

$$
A{B}
\prod_{k=1}^{n-2}\int_0^{\pi}C_{m_k-m_{k+1}}^{m_{k+1}+\frac{n-1}{2}-\frac{1}{2}k}(cos{\theta_{n-k}})
C_{l_k-l_{k+1}}^{l_{k+1}+\frac{n-1}{2}-\frac{1}{2}k}(cos{\theta_{n-k}})[sin{\theta_{n-k}}]^{m_{k+1}+l_{k+1}+n-3}
d{\theta_{n-k}}\times
$$

$$
(A10)\hspace{2.5in} 
\int_0^{2\pi}\frac{e^{i(\pm m_{n-1}\mp l_{n-1})\theta_1}}{sin{\theta_1}}
\hspace{3in}
$$ 
or, in an obvious notation

$$
(A10')\hspace{1.5in}
\int_C d^{n}{\eta}Y_{\alpha}^N(\eta)Y_{\alpha'}^{N'}(\eta) =
A\big{[}aKF(n)\delta_{\alpha \alpha'} + B(n)H(n)\big{]}
\hspace{1.5in}
$$ 
with $ F(4)$ in (A6) emerging from $ F(n)$ at $ n \rightarrow 4$ in the context of a double summation over $ \alpha$ and $ \alpha'$ \cite{T}, with

$$
(A11)~~\hspace{1.0in}
A = \frac{1}{(N'+\frac{n}{2}-1)(N'+\frac{n}{2}) - (N+\frac{n}{2}-1)(N+\frac{n}{2})};~~N \neq N' 
\hspace{1.5in}
$$ 
and with

$$  
{B(4)} = 2\big{[}m_1(sin{\theta_4^0})^{m_{1}+l_{1}+2}(cos{\theta_4^0})
C_{N-m_{1}}^{m_1+\frac{3}{2}}(cos{\theta_4^0})~~ - $$

$$
(A12)\hspace{1.0in}  
(sin{\theta_4^0})^{m_{1}+l_{1}+4}(2m_1+3)C_{N-m_{1}-1}^{m_1+\frac{5}{2}}(cos{\theta_4^0})\big{]}
C_{N'-l_{1}}^{l_1+\frac{3}{2}}(cos{\theta_4^0})
\hspace{3in}
$$

The expression for $ H(4)$ stems from the second term in (A10) and is 

\newpage

$$
H(4) = \prod_{k=1}^{2}\int_0^{\pi}C_{m_k-m_{k+1}}^{m_{k+1}+\frac{3}{2}-\frac{1}{2}k}(cos{\theta_{4-k}})
C_{l_k-l_{k+1}}^{l_{k+1}+\frac{3}{2}-\frac{1}{2}k}(cos{\theta_{4-k}})[sin{\theta_{4-k}}]^{m_{k+1}+l_{k+1}+1}
d{\theta_{4-k}}\times $$

$$
(A13)\hspace{2.0in}
\int_0^{2\pi}\frac{e^{i(\pm m_{3}\mp l_{3})\theta_1}}{sin{\theta_1}}d\theta_1 
\hspace{2.5in}
$$

\end{document}